\documentclass[aps,prx,twocolumn,groupedaddress,superscriptaddress,showpacs,longbibliography]{revtex4-1}

\usepackage{amsfonts,amssymb,amsbsy,amsmath}
\usepackage{graphicx}
\usepackage{epstopdf}
\usepackage{amsthm}
\usepackage{subfigure}
\usepackage{hhline}
\usepackage[miktex]{gnuplottex}

\usepackage{color} % para escribir texto de colores.
\usepackage[normalem]{ulem} % para tachar texto.

\newcommand{\avrg}[1]{\left\langle #1 \right\rangle}

\newcommand{\nn}{\nonumber \\}

\newcommand{\argmax}[1]{\underset{#1}{\mathrm{argmax}}}
\newcommand{\argmin}[1]{\underset{#1}{\mathrm{argmin}}}

\DeclareMathOperator{\Tr}{Tr}
\newcommand{\from}{\leftarrow}

\begin{document}

\title{
Thermodynamics of the Minimum Description Length on Community Detection
}

\author{Juan Ignacio Perotti}
\affiliation{Facultad de Matem\'atica, Astronom\'ia, F\'isica y Computaci\'on, Universidad Nacional de C\'ordoba, Ciudad Universitaria, 5000 C\'ordoba, Argentina}
\affiliation{Instituto de F\'isica Enrique Gaviola (IFEG-CONICET), Ciudad Universitaria, 5000 C\'ordoba, Argentina}
\email[E-mail: ]{juanpool@gmail.com}

\author{Claudio Juan Tessone}
\affiliation{URPP Social Networks, Universit\"at Z\"urich, Andreasstrasse 15, CH-8050 Z\"urich, Switzerland}
\email[E-mail: ]{claudio.tessone@business.uzh.ch}

\author{Aaron Clauset}
\affiliation{Department  of  Computer  Science,  University  of  Colorado,  Boulder  CO,  80309  USA}
\affiliation{Santa  Fe  Institute,  Santa  Fe  NM,  87501  USA}
\affiliation{
BioFrontiers  Institute,  University  of  Colorado,  Boulder  CO,  80303  USA}
\email[E-mail: ]{aaron.clauset@colorado.edu}

\author{Guido Caldarelli} 
\affiliation{IMT School for Advanced Studies Lucca, Piazza San Francesco 19, I-55100, Lucca, Italy}
\affiliation{Institute for Complex Systems CNR, via dei Taurini 19, I-00185, Roma, Italy}
\affiliation{London Institute for Mathematical Sciences, 35a South St. Mayfair, London W1K 2XF UK}
\affiliation{European Centre for Living Technology (ECLT), San Marco 2840, 30124, Venezia, Italy}
\email[E-mail: ]{guido.caldarelli@imtlucca.it}

\date{\today}

\begin{abstract}
%1. 
Modern statistical modeling is an important complement to the more traditional approach of physics where Complex Systems are studied by means of extremely simple idealized models.
%2.
The Minimum Description Length (MDL) is a principled approach to statistical modeling combining Occam's razor with Information Theory for the selection of models providing the most concise descriptions.
%3.
In this work, we introduce the Boltzmannian MDL (BMDL), a formalization of the principle of MDL with a parametric complexity conveniently formulated as the free-energy of an artificial thermodynamic system.
In this way, we leverage on the rich theoretical and technical background of statistical mechanics, to show the crucial importance that phase transitions and other thermodynamic concepts have on the problem of statistical modeling from an information theoretic point of view.
%4.
For example, we provide information theoretic justifications of why a high-temperature series expansion can be used to compute systematic approximations of the BMDL when the formalism is used to model data, and why statistically significant model selections can be identified with ordered phases when the BMDL is used to  {\em model models}.
%5.
To test the introduced formalism, we compute approximations of BMDL for the problem of community detection in complex networks, where we obtain a principled MDL derivation of the Girvan-Newman (GN) modularity and the Zhang-Moore (ZM) community detection method.
%6
Here, by means of analytical estimations and numerical experiments on synthetic and empirical networks, we find that BMDL-based correction terms of the GN modularity improve the quality of the detected communities and we also find an information theoretic justification of why the ZM criterion for estimation of the number of network communities is better than alternative approaches such as the bare minimization of a free energy.
%7
Finally, we discuss several research questions for future works, contemplating the general nature of the BMDL and its application to the particular problem of community detection in complex networks.
\end{abstract}

%89.75.Hc Networks and genealogical trees
%05.45.-a Dynamical systems nonlinear, 
%89.75.-k Complex systems, 
%89.75.Fb Structures and organization in complex systems
%64.60.ah Percolation
%02.50.Ey: Stochastic processes 52,5
%05.45.Tp: Time series analysis
%05.40.-a Fluctuation phenomena, random processes, noise, and Brownian motion
\maketitle

\section{INTRODUCTION}

Physics traditionally emphasizes the use of simple idealized models, an advantageous practice favoring comprehension and facilitating the development of dedicated experiments.
But the emerging information era constantly creates new opportunities for the fruitful use of complex models within the context of statistical modeling.
In this regard, the ultimate goal is to enable a non-trivial statistically significant synthesization of information and theories from the available data.
There are several alternative frameworks for statistical modeling. 
The Bayesian~\cite{von2014bayesian}, the Minimum Message Length (MML)~\cite{wallace2005statistical} and the Minimum Description Length (MDL)~\cite{rissanen1978modeling,
barron1998minimum,
hansen2001model,
grunwald2007minimum} are among the most popular ones.
In a sense, all these methods seek to extract statistically significant information from the hidden patterns laying in the data by by filter the noise through different means.
In particular, frameworks based on the MDL propose to filter noise through data compression, since random patterns are almost always informationally incompressible, combining Occam's razor and information theory to favor models resulting in the shortest descriptions of the data.
In this work, we introduce the Boltzmannian MDL (BMDL), a statistical mechanical formulation of the MDL principle, leveraging on the rich theoretical and technical background already existing for the study o systems in thermodynamic equilibrium~\cite{binney1992theory,stanley1972introduction,
peliti2011statistical}, 
to study the role that fundamental concepts such as phase transitions have from the information theoretic point of view, and to exploit different statistical mechanical approximation methods for the actual computation of BMDL codeword-lengths~\cite{mezard2009information}.

To illustrate how the introduced formalism works on against real problems in statistical modeling, we confront the BMDL framework with the study of divergences in the Refined MDL (RMDL)~\cite{barron1998minimum} and the to study community structures in complex networks~\cite{fortunato2010community}.
In particular, by combining a family of statistical models with the BMDL formalism to model data, we obtain a principled derivation of the Girvan-Newman (GN) modularity~\cite{girvan2002community} with corresponding correction terms which improve the detection performance.
Similarly, by using the BMDL to {\em model models}, we also derive the Zhang-Moore (ZM) Belief Propagation community detection method~\cite{zhang2014scalable}, together with an information theoretic justification of why their criterion to infer the number of network communities works better than other alternatives.
Our results show the flexibility of the BMDL for the particular problem of community detection, enabling several opportunities for future works, but we also remark the general applicability of the BMDL formalism on the broader context of statistical modeling.

In Secs.~\ref{sec:RMDL} different formalizations of the MDL principle are revisited.
The BMDL extension is introduced in Sec.~\ref{sec:betaRMDL} and the high-temperature series expansion is introduced in Sec.~\ref{sec:high_T_series}.
In Secs.~\ref{sec:eff_hamiltonian}~and~\ref{sec:statistical_significance} we discuss information theoretic aspects of statistical significant model selections.
In Sec.~\ref{sec:results:divergences} we show how the BMDL can be used to characterize and fix the emergence of singularities on MDL-based universal codes and in Sec.~\ref{sec:community_detection} we test the formalism in the practical problem of community detection on complex networks.
Finally, our contributions are summarized in Sec.~\ref{sec:conclusions}, where future lines of research are also mentioned.

\section{Theory}

\subsection{Formalizations of the MDL principle}
\label{sec:RMDL}

Consider a {\em statistical model} or probability distribution $P(X)$ defined over some set or configuration space $\mathcal{X}$.
Here $X$ denotes a stochastic variable, $x$ a realization or stochastic variate of $X$ and $P(x)$ a probability.
A parameterized conditional probability distribution $P(X|\theta)$ is called a {\em model} when it is considered as a function of $\theta\in \Theta$.
Each choice of $\theta$  determines a {\em statistical model} $P(X|\theta)$.
A {\em family of models} is a set of models parameterized by some index $m\in\mathcal{M}$ where each member $P_m(X|\theta_m)$ of the family may have its own parameterization domain $\Theta_m$.
The idea is that different members of the family comprehend different modeling choices aiming to represent different patterns of the data with potentially different levels of complexity.

Informally, the principle of Minimum Description Length (MDL) states: {\em Given some data and a family of statistical models, the preferred model provides its most concise description}.
In practice, any formalization of the principle of MDL requires a quantitative definition of the {\em conciseness} of the competing models, which is usually provided using ideas from
Information Theory~\cite{cover2006elements}.
According to C. Shannon,
it is impossible to communicate an arbitrarily large sequence of i.i.d. variates generated from a probability distribution $P(X)$ by using a message with  lesser than 
$\langle L_P\rangle_P 
= 
\sum_x -P(x)\ln P(x)$ 
nats of information per variate~\footnote{Throughout the paper, information is measured in units of nats. One nat equals $1/\ln 2\approx 1.44$ bits.}.
Here $\langle L_P\rangle_P$ is the Shannon's entropy of the distribution $P(X)$ and
$L_P(x)=-\ln P(x)$ is the length in nats of the codeword for $x$ of Shannon's optimal code for the variates in $\mathcal{X}$.
Shannon's result can be exploited to formalize the principle of MDL by mapping the different models $P_m$ into corresponding distributions $Q_m$, so the description length of any variate $x\in \mathcal{X}$ could be defined as $L_m(x):=-\ln Q_m(x)$.
If the mapping is appropriate, then good or bad models $P_m$ should result in relatively short or large description lengths $L_m(x)$, respectively.

Several mappings form models $P_m$ to distributions $Q_m$ there exist.
For example, the two-part MDL---or equivalently, the Bayesian Maximum a Posteriori (MaP)---where
$Q_m(x)
:=
P_m(x|\underline{\theta}_m)
P_m(\underline{\theta}_m)
$,
$P_m(\theta_m)$ is a given prior distribution and
$\underline{\theta}_m$ 
maximizes the posterior
$
P(\theta_m|x)
\propto 
P_m(x|\theta_m)
P_m(\theta_m)
$ for the given $x$.
Another possibility comes from a Bayesian mixture where
$Q_m(x)
:=
\sum_{\theta_m}
P_m(x|\theta_m)
P_m(\theta_m)$.
In particular, we are interested on the mapping defining the so called Refined MDL (RMDL).
At difference with the previous examples, the RMDL is not Bayesian alike and it is defined by
\begin{equation}
\label{eq:NML}
Q_m(x)
:=
\frac{1}{Z_m}
%P_m(x|%\overline{\theta}_m(x))
W_m(x)
\end{equation}
where
\begin{equation}
W_m(x)
:=
P_m(x|\overline{\theta}_m(x))
\end{equation}
is the Maximum Likelihood (ML) associated to $P_m$ at $x$,
\begin{equation}
\overline{\theta}_m(x)
:=
\argmax{\theta_m}
\;
P_m(x|\theta_m)
\end{equation}
is the corresponding Maximum Likelihood Estimator (MLE)
and
\begin{equation}
\label{eq:RMDL_Z}
Z_m
:=
\sum_x
%P_m(x|%\overline{\theta}_m(x))
W_m(x)
\end{equation}
is a normalization constant.
The distribution $Q_m(x)$ in 
Eq.~\ref{eq:NML} is called the Normalized Maximum Likelihood (NML) of model $P_m$ and is motivated by the following results~\cite{barron1998minimum}.
Firstly, notice that $W_m$ is not a distribution because
\begin{equation}
\sum_x
W_m(x)
%=
%\sum_x
%P_m(x|\hat{\theta}_m(x))
>
\sum_x
P_m(x|\theta_m)
=
1
\end{equation}
for any $\theta_m \in \Theta_m$~\footnote{Strictly speaking, the inequality becomes an identity when $P_m(x|\theta_m)$ is a independent function of $\theta_m$. 
But such trivial case is not of interest for us.}.
Hence, the choice $Q_m:=W_m$
is useless since
Kraft's inequality fails~\cite{cover2006elements}
resulting in a non-decodable code and, in consequence, an ill defined formalization of the principle of MDL.
Secondly, the NML is the solution of the following minimax problem~\footnote{Sometimes the minimax approach is critisized. Hence, a similar derivation of the NML in terms of expected values of the regret can be also obtained~\cite{rissanen2001strong}.}
\begin{equation}
Q_m
:=
\argmin{Q:\sum_x Q(x)=1}
R_m(Q)
\end{equation}
where
\begin{equation}
R_m(Q)
:=
\max_x
\big(
-
\ln Q(x)
+
\ln W_m(x)
\big)
\end{equation}
is called the {\em regret}.
The regret $R_m$ is a sort of distance between the codes associated to $Q$ and $W_m$.
Informally, the idea is that the solution $Q_m$ mimics $W_m$ as much as possible without giving away decodability.

The RMDL of $x$---also called the {\em statistical complexity} of $x$---can be split into two terms
\begin{eqnarray}
\label{eq:RMDL}
L_m(x)
&=&
-
\ln Q_m(x)
=
H_m(x)
+
\ln Z_m
.
\end{eqnarray}
Here, the first term
$H_m(x)
:= -\ln W_m(x)$ accounts for how well model $P_m$ is able to fit the variate $x$ and 
the second term---usually called the {\em parametric complexity}~\cite{grunwald2007minimum}---accounts for how well model $P_m$ is able to fit each of the variates in $\mathcal{X}$; a sort of quantification of the flexibility or descriptive power of model $P_m$.
The idea is that simple models cannot fit $x$ well resulting in large values of $H_m(x)$---meaning under-fitting---while complex models fit well almost all variates in $\mathcal{X}$ resulting in large values of $\ln Z_m$---meaning over-fitting.
Good models balance these two extremes providing the shortest description or best data compression for the datum $x$.
Hence, in the RMDL framework, the minimization of $L_m(x)$ with respect to $m\in \mathcal{M}$ for the given $x\in\mathcal{X}$ defines a criterion for model selection.

\subsection{
\label{sec:betaRMDL}
The Boltzmannian MDL (BMDL)
}

The combination of statistical mechanics and statistical modeling is not new~\cite{
hastings2006community,
mastromatteo2011criticality,
zhang2014scalable,
zdeborova2016statistical} 
even in the context of the MDL~\cite{balasubramanian2005mdl,
peixoto2013parsimonious}.
In particular, notice that Eq.~\ref{eq:RMDL_Z} for the RMDL resembles a partition function, but since it has no analogous of a temperature, its associated thermodynamic system may non-flexibly stay at different thermodynamic phases depending on the circumstances.
As we are going to show, the thermodynamic inflexibility of the of the RMDL is not necessarily the most convenient approach, reason for which we introduce the so called Boltzmannian MDL (BMDL).
Formally, the BMDL is introduced by extending the proposed family of models $\mathcal{M}$, by replacing the non-decodable codes of codeword-lengths $H_m(x)$ with implicitly defined new ones of codeword-lengths
\begin{equation}
\label{eq:extended_greedy_codes}
H_{\beta, m}(x)
:=
\beta 
H_m(x)
.
\end{equation}
Then, once the extended family of models denoted by $\beta\mathcal{M}$ is introduced, the standard procedure of section~\ref{sec:RMDL} is subsequently applied to obtain the so called BMDL code of codeword-lengths
\begin{equation}
\label{eq:bRMDL}
L_{\beta, m}(x)
=
\beta H_m(x)
+
\ln
Z_m(\beta)
%\nonumber
\end{equation}
where
\begin{equation}
\label{eq:betaRMDL:partfun}
Z_m(\beta)
=
\sum_x
e^{-\beta H_m(x)}
%\nonumber
\end{equation}
can be recognized as the partition function of a fictitious statistical mechanical thermodynamic system with Hamiltonian $H_m(x)$, inverse temperature $\beta$, Boltzmann distribution $Q_{\beta,m}(x)=e^{-\beta H_m(x)}/Z_m(\beta)$ thermodynamic energy $U_m(\beta):=\sum_x Q_{\beta,m}(x)H_m(x)$, entropy $S_m(\beta)=\sum_x Q_{\beta,m}(x)L_{\beta,m}(x)$ and free energy 
$F_m(\beta)=-\frac{1}{\beta}\ln Z_m(\beta)$.
This clearly establishes the connection between the BMDL formalism and the canonical ensemble of statistical mechanics.

\subsection{
\label{sec:high_T_series}
A high-temperature cumulant expansion for the BMDL
}

For fixed $m$ and as $\beta$ varies, the fictitious thermodynamic system described by the partition function of
Eq.~\ref{eq:betaRMDL:partfun} may undergo 
through different phase transitions.
This greatly complicates the computation, approximation or estimation of the parametric complexity $\ln Z_m$.
Although different methods exist to characterize $\ln Z_m$ at the different phases---e.g. mean field (MF), Bethe-Peierls' approximations~\cite{mezard2009information}, series expansions~\cite{stanley1972introduction}, replica and cavity methods~\cite{fischer1993spin}, Renormalization Group~\cite{binney1992theory} and Monte Carlo methods~\cite{newman1999monte}---the combined use of them is problematic.
Firstly, different methods tend to introduce different biases into the approximations, so a quantitative comparison of $\ln Z_m$ at the different phases becomes unreliable.
Secondly and, importantly, 
most approximation methods somehow reflect the spontaneous emergence of broken symmetries or ergodicity breaking characteristic of the low-temperature phases---even for finite but sufficiently large systems---something that should be avoided to keep the code decodable over the whole of $\mathcal{X}$ or, at least, without the spontaneously emerging statistical biases not originally intended in the definition of $Q_{\beta,m}(X)$.
Hence, to avoid these problems, we propose to restrict the values of $\beta$ to the high-temperature region, and use a high-temperature series expansion for the estimation of $\ln Z_m$.
This approach conveys several advantages:
{\em i)} since the high-temperature regime always exists for any inverse temperature below some critical value, any family of models can be always extended.
{\em ii)} the high-temperature series expansion is usually computationally much cheaper than most other alternatives.
{\em iii)} the interpretation of terms of a high-temperature series expansion gives useful insights on how the BMDL operates for statistical modeling.

Assuming the $m$-th thermodynamic system is being at the high-temperature regime, the high-temperature cumulant expansion reads~\cite{stanley1972introduction}
\begin{eqnarray}
\label{eq:hightemp:series}
\ln Z_m(\beta)
&=&
\ln
\bigg(
\sum_x
e^{-\beta H_m(x)}
\bigg)
\\
&=&
%\ln 
%|\mathcal{X}|
\ln 
Z_0
+
\ln
\bigg\langle
\sum_x
e^{-\beta H_m(x)}
\bigg\rangle_0
\nn
&=&
%\ln |\mathcal{X}|
\ln Z_0
-
\beta
\langle
H_m
\rangle_0
+
...
+
\frac{
(
-\beta
)^{\ell}
}{\ell !}
\langle
(H_m)^{\ell}_c
\rangle_0
+
...
\nonumber
\;.
\end{eqnarray}
Here, $Z_0=|\mathcal{X}|$ is the volume of $\mathcal{X}$,
$
\langle f \rangle_0
=
%|\mathcal{X}|^{-1}
Z_0^{-1}
\sum_x f(x)
$
denotes the infinite temperature average of any function $f(x)$ and
$\langle (f)_c^{\ell}\rangle_0$
denotes the $\ell$-th corresponding cumulant.
The first term of the series
$\ln Z_0$ 
is a constant independent of $(\beta,m)$ and corresponds to the limit $\beta\to 0^+$.
The other terms are cumulants---including the average---each of which is usually easier to compute or estimate than the full of $\ln Z_m$, mainly because they are defined in the limit of infinite temperature.
When $\beta$ approaches the transition point, the high-temperature series expansions may converge slowly.
Otherwise, a few terms of the expansion already provide good approximations.
In fact, four our purposes, it is convenient to consider the approximation provided
by the first three terms of the expansion
\begin{eqnarray}
\label{eq:series_betaRMDL}
L_{\beta, m}(x)
&\approx &
%\ln |\mathcal{X}|
\ln Z_0
+
\beta
\big(
H_m(x)
-
\langle
H_m
\rangle_0
\big)
+
\\
&&
\hspace{.5cm}
+ 
\tfrac{\beta^2}{2}
\big(
\langle
H_m^2
\rangle_0
-
\langle
H_m
\rangle_0^2
\big)
%+
%...
%.
\nonumber
\end{eqnarray}
because
the following reasons.
Besides the constant term 
$\ln Z_0$,
the first order contribution estimates how much the information describing $x$ is compressed by the given code as compared to a corresponding average over the whole of
$\mathcal{X}$.
The second order term---which approximates a heat capacity---can be used to detect the proximity of $\beta$ to the transition points where the high-temperature phase ends and a low-temperature phase begins.
Specifically, since it grows with $\beta$, it automatically 
helps to discard choices of $(\beta,m)$ 
corresponding to thermodynamic systems outside their high-temperature regimes.

\subsection{
\label{sec:eff_hamiltonian}
Effective Hamiltonians
}

In general, 
$\ln Z_0$ 
is a quantity difficult to compute or estimate, but since it is a constant independent of $\beta$ and $m$, it is convenient to introduce the effective Hamiltonian
\begin{eqnarray}
\label{eq:eff_hamilt}
\mathcal{H}_{\beta, m}(x)
&:=&
\tfrac{1}{\beta}
\big(
L_{\beta, m}(x)
-
\ln Z_0
\big)
\\
&=&
H_m(x)
-
\langle
H_m
\rangle_0
+
\tfrac{\beta}{2}
\big(
\langle
H_m^2
\rangle_0
-
\langle
H_m
\rangle_0^2
\big)
+
...
\nn
&=&
\mathcal{H}_m^{(0)}(x)
+
\beta
\mathcal{H}_m^{(1)}
+
...
+
\beta^{\ell}
\mathcal{H}_m^{(\ell)}
+
...
\nonumber
\end{eqnarray}
which, for fixed $\beta$ and $x$, is an objective function of $m$ equivalent to $L_{\beta,m}(x)$.
In particular, since $\mathcal{H}
_{\beta,m}(x)<0
\Leftrightarrow
L_{\beta, m}(x)
<
%\ln |\mathcal{X}|
\ln Z_0
$,
the condition $\mathcal{H}
_{\beta,m}(x)<0$ can be interpreted as a sort of {\em effective compressibility condition} where the $(\beta,m)$-th model is able to compress the information necessary to describe $x$ more than it can be compressed by the null model 
$P_0(x)=Z_0^{-1}$.

\subsection{
\label{sec:statistical_significance}
Statistical significant model selections
}

In a sense, the 
BMDL
formalism is analogous to the Maximum a Posteriori (MaP) Bayesian approach~\cite{von2014bayesian}.
It corresponds to an optimization problem, well suited to select the best fitting
models, but not to judge their statistical significance.
Here, we re-use the BMDL framework but to {\em model models}, in order to introduce a formalism akin to the full Bayesian approach which can be used to judge and find statistical significant model selections.
For this purpose, lets consider $x$ as a fixed quantity and---in analogy to the use of hyper-parameters in Bayesian modeling---we define the so called hyper Hamiltonian $H'_{x}(\beta,m):=\mathcal{H}_{\beta,m}(x)$---a function from $\beta \mathcal{M}$ to the real numbers.
%with the purpose of {\em model models}.
In this way, a corresponding BMDL analogous to that of section~\ref{sec:betaRMDL} can be derived
\begin{equation}
\label{eq:metaRMDL}
L'_{\beta',x}(\beta,m)
:=
\beta' H'_{x}(\beta,m)
+
\ln Z'_{x}(\beta')
\end{equation}
but 
which is meant to represent codeword-lengths for the elements within $\beta\mathcal{M}$.
We speak of $L'$ as the hyper BMDL or, more briefly, the B'MDL.
Its partition function---called hyper partition function---is defined by
\begin{eqnarray}
\label{eq:metaRMDL_Z}
Z'_{x}(\beta')
&=&
\sum_{m\in \mathcal{M}}
\int_0^{\infty}
d\beta \,
P(\beta)
e^{-\beta'H'_{x}(\beta,m)}
\\
&=&
\Tr
e^{-\beta'H'_{x}(\beta,m)}
\nonumber
\end{eqnarray}
and we refer to its associated thermodynamic system as the hyper thermodynamic system with
hyper inverse temperature $\beta'$.
Here, $P(\beta)$ is a probability density function introduced for completeness.
A Dirac delta $\delta(\beta)$ or a step function $\frac{1}{\beta_h}\Theta(\beta_h-\beta)$ are among the simplest choices for $P(\beta)$.

Informally, a statistically significant model selection in $\beta\mathcal{M}$ can be associated to a deep minimum of $H'_{x}(\beta,m)$ which also has a large basin of attraction that clearly favors certain models.
From the practical point of view, the question is how to formally characterize the existence of such minima.
For finite systems or at the high-temperature regime, 
$H'_x(\beta,m)$ 
and
$L'_{\beta',x}(\beta,m)$ 
are equivalent objective functions of $(\beta,m)$
since $Z_x(\beta')$ behaves like a constant.
However, for infinite---or in practice, sufficiently large---systems, the existence of phase transitions and the accompanying ergodicity breaking creates a dependency of $Z'_x(\beta')$ with respect to $(\beta,m)$, disrupting the equivalence between $H'_x(\beta,m)$ 
and $L'_{\beta',x}(\beta,m)$.
In fact, in a low-temperature phase, $\beta\mathcal{M}$ breaks into several thermodynamic basins of attraction and, correspondingly, the information theoretic code defined by the B'MDL split into several codes, one for each basin or model selection.
Hence, in the low-temperature phases, the value of $L'_{\beta',x}(\beta,m)$ is rendered unusable for model selection over the whole of $\beta\mathcal{M}$ but it still works on each of the basins separately.
In this regard, the choice of one of the basins based on its {\em thermo-statistical attractiveness} can be interpreted as a principled statistically significant model selection.
The rationale of considering ordered phases as the only representative of significant selections, is that ordered phases correspond to the only {\em macroscopically} large thermo-statistical basins of attraction covering non-negligible fractions of $\beta\mathcal{M}$.
Here, we had followed the ideas in~\cite{hastings2006community,
decelle2011inference,
zhang2014scalable}, 
but we added an information-theoretic justification for the criterion.

Let us remark some important differences between the BMDL and the B'MDL formalisms.
The former is devised 
to comparatively weigh the goodness of fit of  different models, it is assumed to be used on the high-temperature regime and it corresponds to an optimization problem analogous to the MaP Bayesian approach.
The later is devised to find statistical significant model selections,
it is meant to work at ordered phases and it essentially poses an integral problem similar to the full Bayesian approach.
Besides these differences, notice that the BMDL formalism can be recovered from the zero hyper temperature limit $\beta'\to \infty$ of the 
B'MDL formalism, although the resulting ground state may not correspond to an ordered state or, in other words, to a statistical significant 
model selection.

\section{
\label{sec:results}
Results
}

In what follows we illustrate how the introduced framework work in practice in a couple of examples.
Firstly, we apply the BMDL to a family of geometric distributions to show how the emergence of problematic divergences arising for the RMDL formalism, can be analyzed and eventually treated with the help of the BMDL framework.
Secondly, we apply both, the BMDL
and the B'MDL formalisms to the challenging problem of community detection in complex networks.

\subsection{
\label{sec:results:divergences}
Renormalization of divergences
}

Consider a family of multivariate geometric distributions used to model the measurement of $k_1,...,k_n$ i.i.d. non-negative integer numbers~\cite{rooij2006empirical}.
The $m$-th member of the family is defined to be
\begin{equation}
P_m(k)
=
\prod_{i=1}^n
m^{k_i}(1+m)^{-1-k_i}
\end{equation}
Hence, its Hamiltonian is
(see Eq.~\ref{eq:RMDL})
\begin{equation}
H_m(k_1,...,k_n)
=
\sum_{i=1}^n
h_m(k_i)
\end{equation}
where
\begin{equation}
h_m(k)
=
(1+k)\ln(1+m)
-
k\ln m
\end{equation}
is the contribution per degree of freedom.
We can understand the divergences of the RMDL for the multivariate geometric distribution by studying the high-temperature series expansion of Eq.~\ref{eq:hightemp:series}
for the BMDL
evaluated at $\beta=1$.
Since the configuration space for the multivariate geometric distribution is 
$\mathcal{X}=\mathbb{N}_0^n$, where
$\mathbb{N}_0=\{0,1,2,...\}$ is the set of non-negative integer numbers, then $\mathcal{X}$ is
clearly unbounded. 
Hence, we can identify 
$\ln Z_0$ 
as one of the divergent contributions affecting the RMDL.
Such divergence can be easily cured by the introduction of the effective Hamiltonian of  Eq.~\ref{eq:eff_hamilt} which
is a sort of renormalization procedure.
But this is not the only source of divergent contributions to the BMDL.
The unbounded number of energy levels per degree of freedom characterized by the unboundedness of $h_m$, results in a divergent infinite-temperature average $\avrg{H_m}_0$ of the Hamiltonian.
This sort of divergences---which somehow corresponds to ultra-violet divergences---may also be cured using more sophisticate renormalization procedures
or series expansions~\cite{kardar2007statistical}, which are not discussed here.
Alternatively, as our framework makes 
clear, such divergences can be cured by choosing a more convenient null model.
For instance, for the present case of the multivariate geometric distributions, we can restrict $\mathcal{X}$ to those values of $k_1,...,k_n$ satisfying the condition $\sum_i k_i=K$ where $K$ is a constant measured from the data. 
In such case $|\mathcal{X}|<\infty$ and the RMDL results finite.

\subsection{
\label{sec:community_detection}
Community detection in complex networks
}

The principle of MDL has been already used for for the problem of community detection in complex networks~\cite{fortunato2010community,
rosvall2007information,
rosvall2008maps,
peixoto2013parsimonious,
peixoto2014hierarchical}.
Here, we follow similar steps but within the context of the BMDL formalisms.

\subsubsection{\label{sec:EDM}The Family of External Degree Models (EDM)}

For the purpose of community detection via the 
BMDL, an appropriate family of statistical models is required.
We choose to use a particular simple one, which we called the family of External Degree Models (EDM).
Before continuing, the reader should keep in mind that, besides the family of EDM that we are going to introduce, many other possible modeling choices for community detection do exists.
In particular, it is possible to use for the definition of $H_{\mathcal{P}}(G)$ any already existing community detection score.

Consider some graph or network $G=(V,E)$ of $N$ nodes $i\in V$, $M$ links in $ij=ji\in E$, no self-links and adjacency matrix of entries $G_{ij}=G_{ji}\in\{0,1\}$.
Consider also a set of 
node
labels denoted by $\mathcal{P}=\{p_1,...,p_N\}$ where $p_i$ denotes the label of node $i$.
$\mathcal{P}$ is used to
represent a potential community structure the network may have.
We define the family of EDM by means of corresponding Hamiltonians, where
\begin{eqnarray}
\label{eq:EDM_hamiltonian}
H_{\mathcal{P}}(G)
:=
\sum_{i\in V} q_{i|p_i}
=
K
-
\sum_{i\in V} k_{i|p_i}
\geq 0
\end{eqnarray}
corresponds to the $\mathcal{P}$-th member of the family.
Here,
$q_{i|p}
:=
k_i
-
k_{i|p}$
is called the external-degree of
$i$ with respect to $p$, and it accounts for the number of links connecting $i$ with other nodes not in $p$.
Similarly, 
$K:=\sum_{i\in V} k_i=2M$ is the total degree of $G$ where $k_i$ is the degree of node $i$ and $k_{i|p}:=\sum_{j\in V} G_{ij}\delta_{pp_j}$ is the internal-degree of $i$ with respect to $p$ accounting for the number of links connecting $i$ with other nodes in $p$.
It is worth noting that $H_{\mathcal{P}}(G)$ equals twice the number of inter-community links of $G$ as determined by $\mathcal{P}$.

\subsubsection{
\label{sec:betaRMDL_EDM}
The BMDL-EDM
method for community detection
}

According to the BMDL formalism, the first order approximation of the high-temperature expansion of the effective Hamiltonian for the family of EDM becomes
\begin{eqnarray}
\label{eq:betaRMDL_EDM_eff_H}
\mathcal{H}_{\beta,\mathcal{P}}
&=&
\big(
H_{\mathcal{P}}
-
\avrg{
H_{\mathcal{P}}
}_0
\big)
+
\tfrac{\beta}{2}
\big(
\avrg{
H_{\mathcal{P}}^2
}_0
-
\avrg{
H_{\mathcal{P}}
}_0^2
\big)
+
...
\nn
&=&
\mathcal{H}_{\mathcal{P}}^{(0)}
+
\beta
\mathcal{H}_{\mathcal{P}}^{(1)}
+
...
\;
.
\end{eqnarray}
The zeroth-order contribution 
$\mathcal{H}_{\mathcal{P}}^{(0)}$
corresponds to the minimization of the total number of inter-community links as compared to its average, while the first-order 
$\beta\mathcal{H}_{\mathcal{P}}^{(1)}$
quantifies the corresponding statistical fluctuations.
Here, averages should be computed over some network ensemble composed of appropriate randomizations of the original network $G$.
In our work, we consider the ensemble of degree preserving random networks~\cite{karrer2011stochastic,
squartini2011analytical}.

A bare minimization of the number of inter-community links $\tfrac{1}{2}H_{\mathcal{P}}(G)$ as a function of the proposed community structure $\mathcal{P}$ leads to the trivial solution composed of one community only $|\mathcal{P}|=1$.
To avoid this problem, the minimization should be compensated by the appropriate contribution of a null model.
This is the underlying idea behind the definition of the Girvan-Newman (GN) modularity~\cite{girvan2002community}. 
In fact, in the limit
$\beta\to 0^+$, the BMDL-EDM---our short name for the combination of the BMDL formalism with the family of EDMs---is reduced to the term
$\mathcal{H}_{\mathcal{P}}^{(0)}(G)$ whose minimization is essentially equivalent to the maximization of the GN modularity.
In other words, we have derived the GN modularity as the zeroth order approximation of the effective Hamiltonian of the BMDL-EDM.
Strictly speaking, GN uses a particular estimation of $\avrg{H_{\mathcal{P}}}_0$ which in the case of the BMDL is still unspecified.
Other derivations of the GN modularity from different statistical modeling frameworks do also exist.
For example, a variant of the GN modularity can be derived from the maximum likelihood approach to statistical modelling combined with the degree corrected stochastic block model~\cite{newman2016equivalence}. 
Ours is the first obtained from the BMDL and the RMDL formalisms.

The high-temperature series expansion greatly simplifies the problem of approximating the 
BMDL, but it does not completely solves the problem.
Problem-specific calculations of the different moments in the expansion are necessary.
Hence, here we introduce appropriate estimations of the moments within
$\mathcal{H}^{(0)}_{\mathcal{P}}$ 
and 
$\mathcal{H}^{(1)}_{\mathcal{P}}$ 
in Eq.~\ref{eq:betaRMDL_EDM_eff_H} for the particular ensemble of degree-preserving random networks.
For alternative ways see~\cite{girvan2002community,traag2015detecting}.
To approximate the zeroth-order effective Hamiltonian, we notice first the identity
$\avrg{H_{\mathcal{P}}}_0
%=
%\avrg{\sum_{i\in V} q_{i|p_i}}_0 
= 
\sum_{i\in V} \avrg{q_{i|p_i}}_0$
so we can approximate the average external degrees by
\begin{eqnarray}
\label{eq:avrg_external_degree}
\avrg{
q_{i|p_i}
}_0
&=&
\avrg{
k_i
}_0
-
\avrg{
k_{i|p_i}
}_0
\approx
\frac{
K
-
K_{p_i}
}{
K
- 
k_i
}
k_i
\end{eqnarray}
where $\avrg{k_i}_0=k_i$,
$K_p
:=
\sum_{j\in V} k_j 
\delta_{p_jp}
$
is the total degree of community $p$
and the quantity
$\frac{
K_{p_i}
- 
k_i
}{
K
- 
k_i
}$ 
approximates the probability for each of the $k_i$ links of $i$ to be connected to other nodes in $p_i$. 
The approximation is consistent with $G$ being considered a multi-graph without self-links.
This approximation improves over the standard analogous approximation considered in the definition of the GN modularity, since the later is proportional to terms of the form $k_ik_j/K$ and may incorrectly approximate probabilities by numbers larger than one when the condition $\sqrt{k_i},\sqrt{k_j} \ll N$ is not satisfied.
Putting all together, the proposed 
approximation for the zeroth-order effective Hamiltonian reads
\begin{eqnarray}
\label{eq:betaRMDL_EDM_eff_H0_estim}
\mathcal{H}_{\mathcal{P}}^{(0)}
&\approx &
H_{\mathcal{P}}
-
\sum_{i\in V}
\frac{K-K_{p_i}}{K-k_i}
k_i
.
\end{eqnarray}
To approximate the contribution of the first-order effective Hamiltonian we notice that it can be written as
$
%\begin{eqnarray}
%\label{eq:betaRMDL_EDM_eff_H1_estim}
\mathcal{H}_{\mathcal{P}}^{(1)}
%&=&
%2\big(
%\avrg{
%\mathtt{E}_{\mathcal{P}}^2
%}_0
%-
%\avrg{
%\mathtt{E}_{\mathcal{P}}
%}_0^2
%\big)
%\nonumber
%\\
%&=&
%2
%\sigma^2_{\mathtt{E}_{\mathcal{P}}}
%\nn
%&=&
=
\tfrac{1}{2}
\sigma^2_{H_{\mathcal{P}}}
$.
In other words, it essentially takes the form of the variance of the total number of inter-community links.
The exact computation of the variance is non-trivial but, by following the ideas used to justify the approximation of Eq.~\ref{eq:avrg_external_degree}, 
it is possible to think the different $k_{i|p_i}$ to be approximately determined by independent binomial processes of $k_i$ events and success probabilities 
$s_i 
= 
\tfrac{
K_{p_i}
-
k_i
}{
K
-
k_i}
$. 
Then, the variance can be approximated by
\begin{equation}
%\begin{eqnarray}
\label{eq:betaRMDL_EDM_eff_H1_estim}
\mathcal{H}_{\mathcal{P}}^{(1)}
%&=&
%2 
%\sigma^2_{\mathtt{E}_{\mathcal{P}}}
%\nonumber
%\\
%&\approx &
\approx
%2
%\frac{1}{2^2}
\frac{1}{2}
\sum_{i\in V} \sigma^2_{k_{i|p_i}}
%\nn
%&\approx &
\approx
\frac{1}{2}
\sum_{i\in V} s_i(1-s_i)k_i
%\nn
%&=&
%\frac{1}{2}
%\sum_{i\in V}
%\frac{
%(
%K
%-
%K_{p_i}
%)(
%K_{p_i}
%-
%k_i
%)
%}{
%(
%K
%-
%k_i
%)^2
%}k_i
%\nn
%&=:&
%\sum_{i\in \partial G}
%\mathcal{H}_i^{(1)}
.
%\nonumber
%\end{eqnarray}
\end{equation}

It is useful to analyze how the approximations
in Eqs.~\ref{eq:betaRMDL_EDM_eff_H0_estim}~and~\ref{eq:betaRMDL_EDM_eff_H1_estim} contribute to $\mathcal{H}_{\beta,\mathcal{P}}(G)$ for the limiting cases of trivial partitions.
In the case of a partition of one community for all nodes where $\mathcal{P}=\{p\}$, it holds that $K_{p}=K$ and therefore 
$\mathcal{H}_{\mathcal{P}}^{(0)}=0$ and
$\mathcal{H}_{\mathcal{P}}^{(1)}=0$.
On the other hand, in the partition of one community per node it holds that $K_{p_i}=k_i$ and, again, both contributions to the effective Hamiltonian equals zero.
In other words, the introduced approximations of the effective Hamiltonian work as they should in the limiting cases of trivial partitions.
They also work correctly in the extreme cases of networks composed of isolated nodes or fully connected networks.
This qualitatively correct behavior of the approximations in the extreme cases is important, since it helps to minimize the emergence of distortive effects.

\subsubsection{
\label{sec:detect_tran}
Detectability Transition
}

We begin characterizing the BMDL as a method for community detection studying its {\em detectability transition}~\cite{decelle2011inference,
mossel2015reconstruction}. 
For this, we use bi-modular synthetic networks whose communities are random graphs with $N_c$ nodes and an expected number of $M_c$ links, connected by a {\em bridge} of an expected number of $M_b$ links.
These networks correspond to the Planted Partition network ensemble, which is a special case of the so called Stochastic Block Model~\cite{fortunato2010community,
decelle2011inference}.
The networks have a total $N=2N_c$ nodes and an expected number of $M=2M_c+M_b$ links.
For $N\gg 1$, the expected link-density is
$\nu 
= 
%\tfrac{M}{\tfrac{N(N-1)}{2}}
\tfrac{M}{N(N-1)/2}
\approx
\tfrac{K}{N^2}$
and the expected link-density of the bridge is
$\nu_b 
=
\tfrac{M_b}{N_c^2}
=
\tfrac{4M_b}{N^2}$.
Ideally, planted partitions should be detectable whenever $\nu-\nu_b>0$.
In practice, however, quenched fluctuations in the structure of the networks make detection harder and the planted partitions can only be detected after some non-negligible value of the difference $\nu-\nu_b>0$.

Consider $\nu$ fixed. We want to work out a simple analytical approximations for the critical value $\nu_b^*>0$ that separates the detectable ($\nu_b<\nu_b^*$) from the non-detectable ($\nu_b>\nu_b^*$) regime.
An estimate for the transition point $\nu_b^*$ can be obtained by demanding the upper bound 
$\mathcal{H}_{\beta,\mathcal{P}_{\mathrm{p.p.}}}(G) = 0$
of the effective compressibility condition (see section~\ref{sec:eff_hamiltonian}) at the planted partition denoted by $\mathcal{P}_{\mathrm{p.p.}}$.
But first, we need to approximate the effective Hamiltonian.
For the bi-modular networks
$\mathcal{H}_{\mathcal{P}_{\mathrm{p.p.}}}^{(0)}(G)
\approx 
2M_b
-
M
%\sum_{p\in \mathcal{P}_{\mathrm{p.p.}}}
\sum_{p}
\sum_{i\in p}
\tfrac{k_i}{K-k_i}$ 
where the summatory can be approximated by
\begin{eqnarray}
\label{eq:summatory_expansion}
\sum_{i\in p}
\frac{k_i}{K-k_i}
&\approx &
N_c
\int dk\,
P_{p}(k)
\frac{k}{K}
\frac{1}{1-\frac{k}{K}}
%\nn
\\
& = &
N_c
\frac{\avrg{k}}{K}
\bigg(
1
+
\frac{1}{K}\frac{\avrg{k^2}}{\avrg{k}}
+
...
\bigg)
\nonumber
.
\end{eqnarray}
Here, $P_p(k)$ represents the degree distribution of the nodes in community $p$ as seen from the whole network and where $\avrg{y^2}\ll \avrg{y}$ for $y=k/K$ is assumed.
Then, after ignoring the high-order terms of the approximation, we find
$\mathcal{H}_{\mathcal{P_{\mathrm{p.p.}}}}^{(0)}
\approx 
M_b-2M_c$.
Similar approximation tricks lead to
$\mathcal{H}_{\mathcal{P}_{p.p.}}^{(1)}
\approx
\tfrac{M}{4}
$.
After joining results, we obtain the following estimate for the effective Hamiltonian
\begin{eqnarray}
\label{eq:estimation_of_H_eff}
\mathcal{H}_{\beta,\mathcal{P}_{\mathrm{p.p.}}}(G)
\approx
\tfrac{N^2}{2}
\big(
\nu_b
-
\nu
+
\tfrac{1}{4}
\beta 
\nu
+
...
\big)
.
\end{eqnarray}
Finally, combining 
this result with the effective compressibility condition, the transition point is found to be %approximately given by 
\begin{equation}
\label{eq:nu_b_crit}
\nu_b^*
\approx
\nu
-
\tfrac{1}{4}
\beta
\tfrac{\avrg{k}}{N}
.
\end{equation}
According to this approximation, the transition point decreases with $\beta$ meaning that quenched fluctuations in the structure of the networks tend to make detection harder.
Moreover, for fixed $\avrg{k}\ll N$ and $\beta>0$, the difference between $\nu$ and $\nu_b^*$ follows the finite-size scaling $\sim 1/N$, predicting that in the thermodynamic limit, the idealized condition $\nu^b=\nu$ is actually realized.

Lets compare the previous analytical estimations with corresponding numerical simulations.
The simulations require the numerical minimization of the effective Hamiltonian of Eq.~\ref{eq:betaRMDL_EDM_eff_H}.
We use an algorithm which is similar to the well known Louvain method for community detection~\cite{blondel2008fast} although is significantly slower since it requires the computation of $\mathcal{H}^{(2)}_{\mathcal{P}}$.
For details on the algorithm please check the source code~\cite{code_RMDL_EDM}.

\begin{figure*}
\mbox{\includegraphics*[width=.31\linewidth]{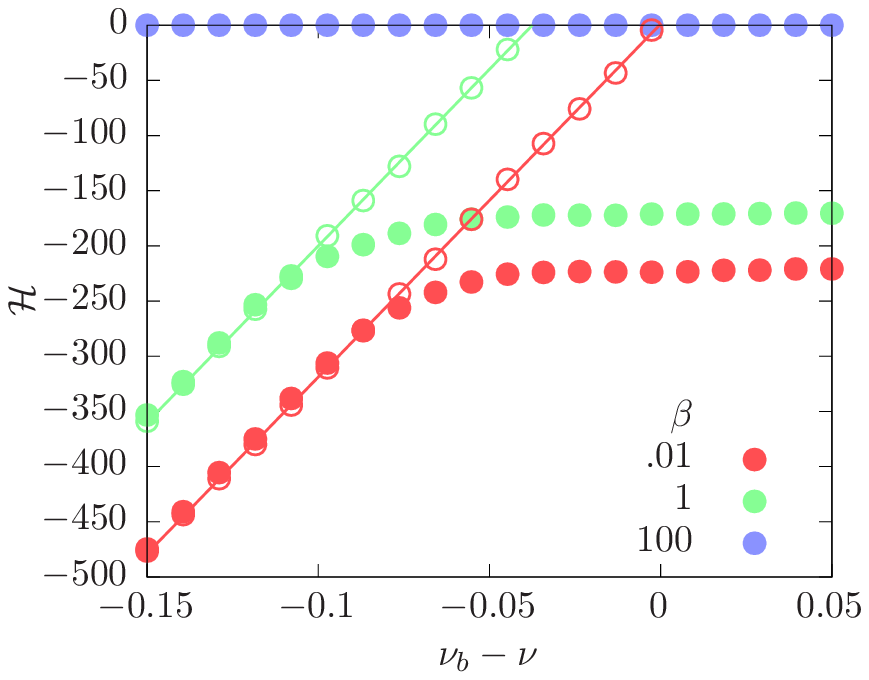}}
\hfill 
\mbox{\includegraphics*[width=.31\linewidth]{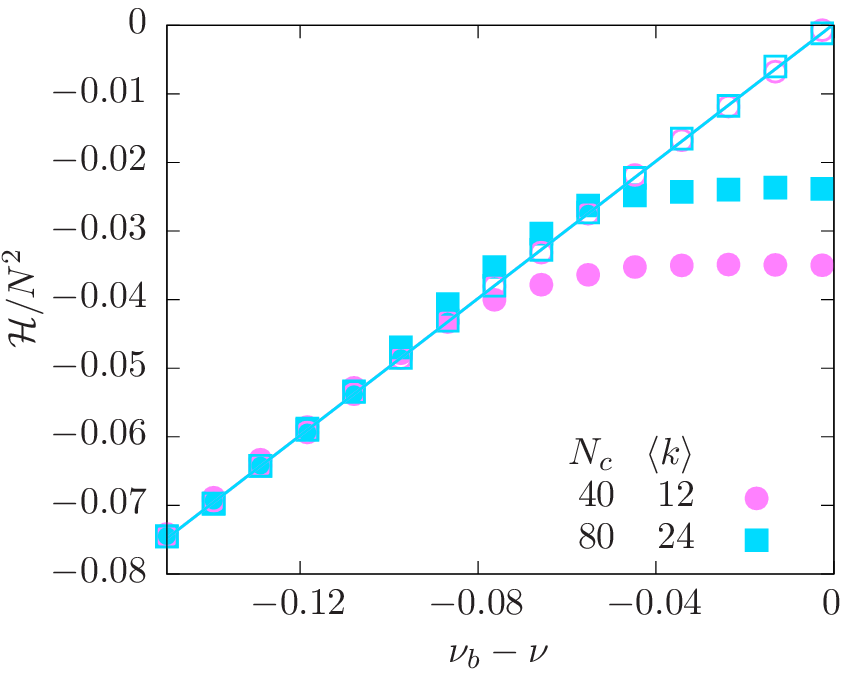}}
\hfill
\mbox{\includegraphics*[width=.31\linewidth]{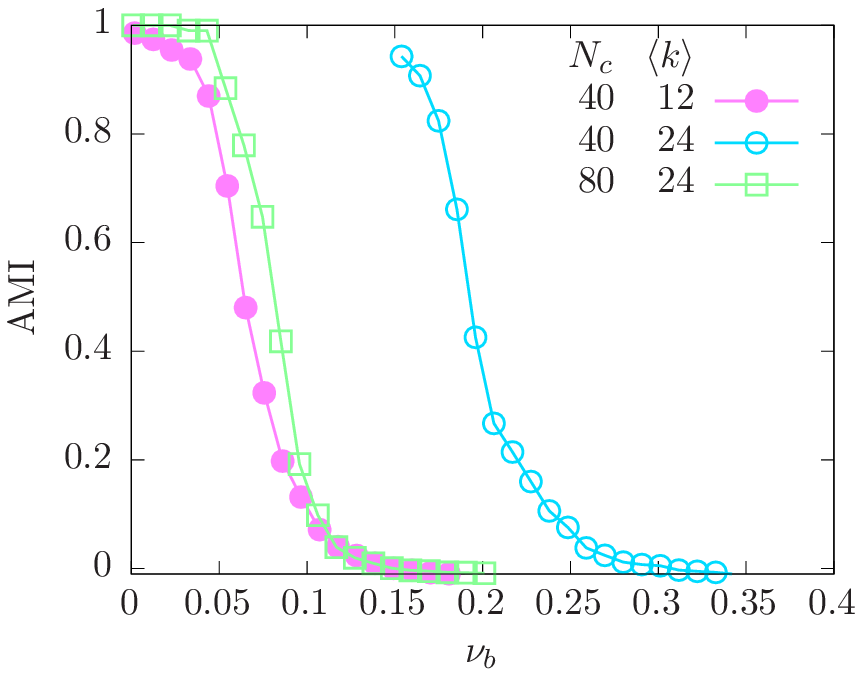}}
\put(-505,120) {a)}
\put(-335,120) {b)}
\put(-160,120) {c)}
\caption{
(Color Online).
Detectability transition for the BMDL-EDM on bi-modular networks of $N_c$ nodes per community and average degree $\avrg{k}$.
In (a), the effective Hamiltonian is plotted as a function of the difference between the inter-community link densities $\nu_b$ and the network link density $\nu$ for networks with $N_c=40$ and $\avrg{k}=12$.
Different inverse temperatures (different colors) for the numerically obtained solutions (solid circles) and the planted partitions (open circles) are shown.
Solid symbols correspond to the numerical solutions and open symbols to the planted partitions.
The lines correspond to the analytical estimation of Eq.~\ref{eq:estimation_of_H_eff}.
Similar curves are shown in (b), but for $\beta=0.01$ and varying values of $N_c$ and $\langle k\rangle$, and a rescaled version of effective Hamiltonian.
In panel (c), the AMI is plotted as a function of $\nu_b$ comparing the numerical obtained partitions with the planted ones. 
Here, different values of $N_c$ and $\avrg{k}$ are considered.
In this and following figures, error bars are negligible compared to symbol sizes and thus are omitted to avoid visual congestion.
}
\label{figTMP:1}
\end{figure*}

In Fig.~\ref{figTMP:1}a, the effective Hamiltonian is plotted as a function of the difference of link-densities $\nu_b-\nu$.
Different colors represent different inverse temperatures.
Open circles correspond to planted partitions, solid circles represent the partitions found by numerical optimization and solid lines represent the estimation of Eq.~\ref{eq:estimation_of_H_eff}.
At low-temperatures (blue), the numerical computations find the trivial solutions of one community per node where no information compression is achieved.
Planted partitions result in a positive effective Hamiltonian in this regime (not shown).
Compression is possible for intermediate- (green) and high-temperatures (red), where the numerical solutions match the planted partitions for sufficiently negative  values of $\nu_b-\nu$.
Eventually, when $\nu_b-\nu$ approaches zero, the planted partitions become suboptimal and a gap emerges between the values of the effective Hamiltonian for the planted partitions and the values for the numerical solutions.
This introduces, a systematic error in the prediction of $\nu_b^*$ of Eq.~\ref{eq:nu_b_crit}.

Fig.~\ref{figTMP:1}b is similar to Fig.~\ref{figTMP:1}a, but a rescaled effective Hamiltonian is shown.
Here, all curves are obtained for $\beta=0.01$ and the different colors (or symbols) represent different choices for $N_c$ and $\avrg{k}$.
As before, solid symbols represent numerical solutions, open symbols the planted partitions and the lines (which overlap) the analytical estimations of Eq.~\ref{eq:estimation_of_H_eff}.
The same gap of Fig.~\ref{figTMP:1}a is observed once $\nu_b-\nu$ approaches zero, but it stretches as $N$ grows in agreement with a finite size prediction of the thermodynamic limit.

In Fig.~\ref{figTMP:1}c the similarity between the detected community structures and the planted partition is quantified by the Adjusted Mutual Information (AMI)---an analogous of the Normalized Mutual Information that compensates for finite size random fluctuations.
As it can be seen, halving $\avrg{k}$ or doubling $N$ has approximately the same effect, which is a prediction of the analytical estimation of Eq.~\ref{eq:nu_b_crit}.

\subsubsection{
Numerical Results on the LFR Benchmark and on empirical networks with the BMDL-EDM
\label{sec:LFR}
}

We systematically test the performance of the 
BMDL-EDM as a method for community detection 
using the LFR benchmark~\cite{lancichinetti2008benchmark}.
This is composed of network ensembles with power-law degree and community size distributions that, to a certain extent, mimic the topology of real networks. 
Each ensemble of the benchmark is generated for a particular value of the so called {\em mixing parameter} $\mu$, whose value determines the difficulty of the community detection problem.
For small $\mu$ the networks have relatively few links between the communities and thus are easy to detect, while as $\mu$ increases the borders between communities blur and the detection problem becomes hard or even impossible to solve.
At $\mu=1$ the intra- and the inter-density of links become equal, the communities stop existing and every community detection method surely fails.
In fact,
in practical conditions all methods surely fail at even smaller values of $\mu$ below some critical value  $\mu_c<1$~\cite{fortunato2010community} in consistency with more rigorous proof with the stochastic block model~\cite{abbe2015community}.

\begin{figure*}
\mbox{\includegraphics*[width=.31\linewidth]{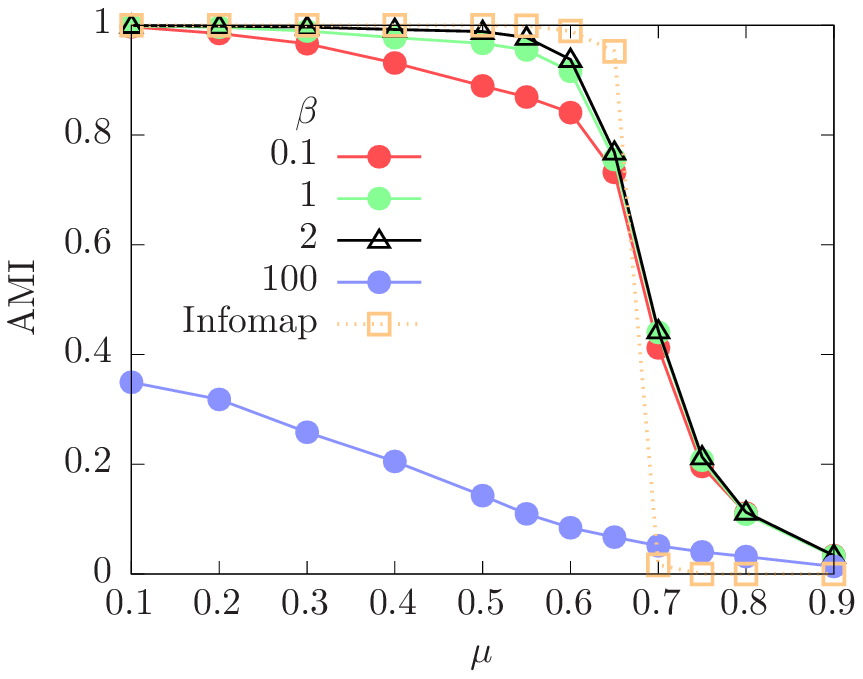}}
\hfill 
\mbox{\includegraphics*[width=.31\linewidth]{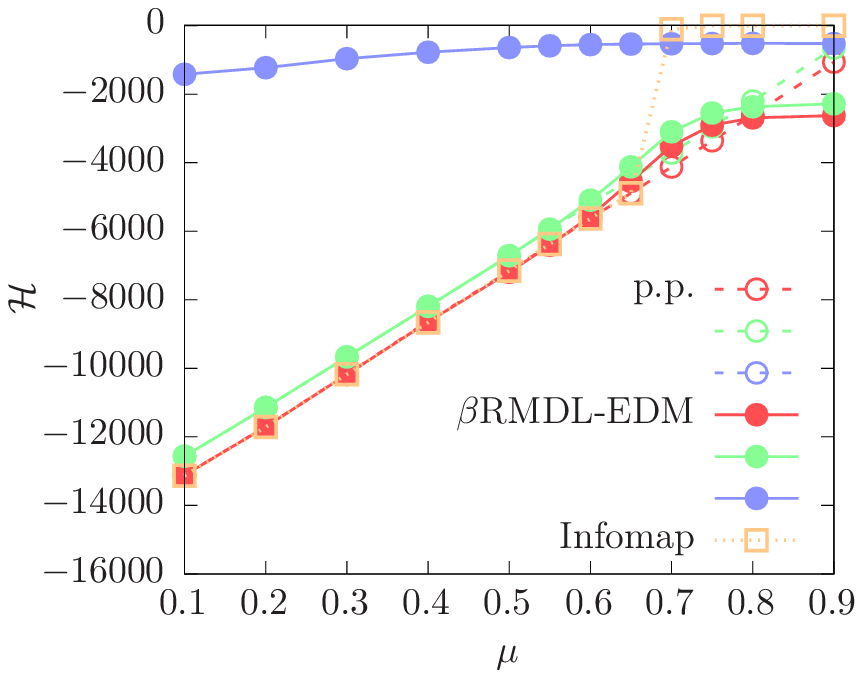}}
\hfill
\mbox{\includegraphics*[width=.31\linewidth]{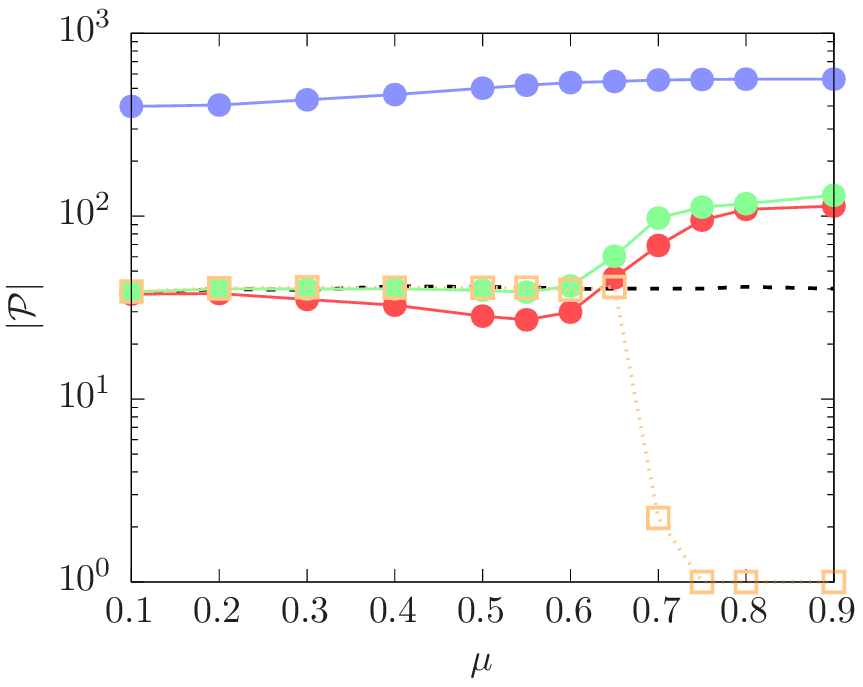}}
\put(-505,120) {a)}
\put(-335,120) {b)}
\put(-160,120) {c)}
\caption{
(Color Online).
LFR benchmark for the BMDL-EDM and for Infomap.
The LFR networks used for the calculations have $N=1000$ nodes, community sizes in the range $[10,50]$ and average degree $\avrg{k}=15$.
In (a), the AMI quantifies the similarity between planted partitions at the different values of $\mu$ and the numerically detected community structures with the BMDL-EDM
(solid circles and open triangles) and Infomap (open squares).
The colors red, green, black and blue correspond to different values of the inverse temperature $\beta$.
In panel (b), the effective Hamiltonian is plotted as a function of the mixing parameter.
The solid circles correspond to the numerical solutions obtained with the BMDL-EDM
open circles to planted partitions and open squares to the community structures detected by Infomap.
In the case of Infomap, the effective Hamiltonian is computed for $\beta=0.1$.
Panel (c) depict curves for the number of detected communities. 
The black dashed line corresponds to the planted number of communities.
}
\label{fig:2}
\end{figure*}

In Fig.~\ref{fig:2} we show results for the BMDL-EDM method over the LFR benchmark.
The numerical solutions were obtained using the same approach of Sec.~\ref{sec:detect_tran}.
In panel~\ref{fig:2}a, the AMI compares the similarity between the planted partitions and the detected community structures.
Solid circles of different colors and the black open triangles correspond to numerical solutions.
The high-, intermediate- and low- temperature regimes are shown in red, green and blue, respectively.
Good results are obtained at high-temperatures, but this is expected because this is the regime where the BMDL-EDM
mimics the GN modularity.
The best results are obtained at some critical temperature $\beta\approx 2$ (black open triangles).
A bad performance is seen at the low-temperature regime.
For comparison, we also show a curve obtained with Infomap~\cite{rosvall2008maps} (open squares).
The comparison suggest that the BMDL-EDM performs as well as the state of the art optimization-based  methods~\cite{fortunato2010community,
yang2016comparative}.

In panel~\ref{fig:2}b, the effective Hamiltonian is plotted as a function of the mixing parameter for both, the numerically obtained community structures (solid circles) and the planted partitions (open circles).
As before, colors correspond to different inverse temperatures and regimes.
Close to the critical value of the mixing parameter $\mu_c\approx .7$  and for the intermediate- (green) and high-temperature (red) regimes, the numeric solutions become sub-optimal since the planted partitions become hard to detect~\cite{decelle2011inference}.
The effective Hamiltonian significantly grows at the low-temperature regime, so some of the curves lay above zero outside the plotted region corresponding to the failure of an effective data compression.
For comparison, values of the effective Hamiltonian at the high-temperature regime are plotted for the communities found by Infomap (open squares) which stops detecting the planted partitions once the hard regime begins.

In panel~\ref{fig:2}c, the detected number of communities is plotted as a function of $\mu$ for the different temperature regimes.
At the intermediate- and hight-temperature regimes, and before the critical value of the mixing parameter, the detected number of communities (solid green and red circles, respectively) match closely the values of the planted partitions (black dashed line).
The same holds for Infomap (open squares).
For values of $\mu>\mu_c$, the numerically optimized BMDL-EDM
method tends to over-estimate the number of communities while Infomap tends to under-estimate it.
This difference in behavior is the reason for which an appropriate comparison between these methods is better
performed with the AMI instead of the traditional NMI~\cite{fortunato2010community}.
This tendency of the BMDL-EDM
method to over-estimate $|\mathcal{P}|$ is a sort of over-fitting; a finding that is consistent with the need of a way to gauge the statistical significance of different selections~\cite{ghasemian2018evaluating},
which we later cover with the combination of the B'MDL formalism and the EDM
(see Sec.~\ref{sec:statistical_significance} and the analysis of its performance in Sec.~\ref{sec:SBP_RMDL_EDM}).

\begin{figure*}
\mbox{\includegraphics*[width=.45
\linewidth,height=5cm,keepaspectratio]{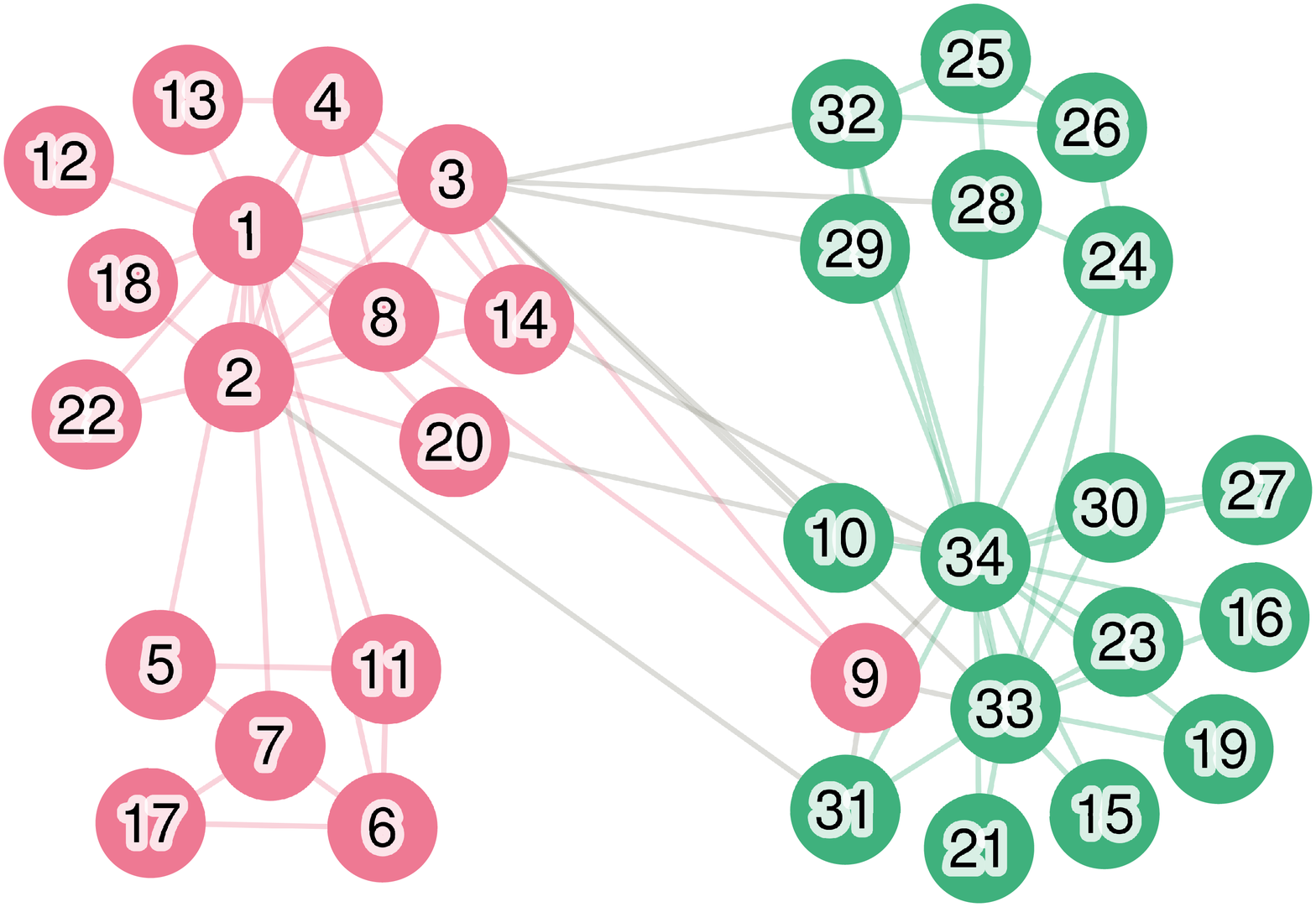}}
\put(-215,125) {a)}
%\\
\hspace{1cm}
\mbox{\includegraphics*[width=.4
\linewidth]{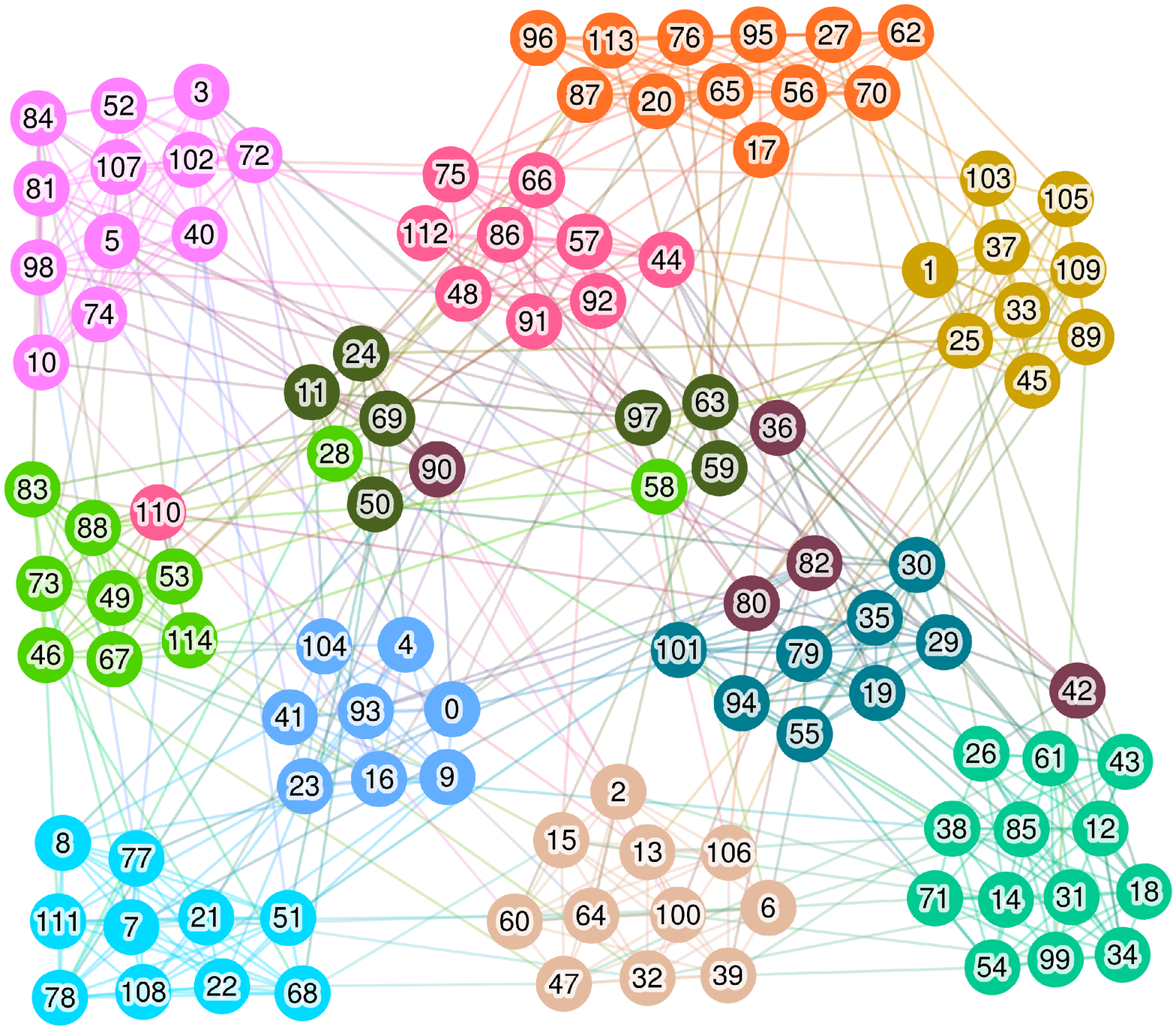}}
\put(-240,125) {b)}
\caption{
(Color Online).
Community structure for
(a) Zachary's karate-club network and (b) American college football, as detected by the minimization of the BMDL-EDM
at $\beta=0$ and $\beta=2$, respectively.
The tight groups of nodes represent the detected community structures while the colors those suggested by the meta-data.
The networks were plotted with Gephi~\cite{gephi}.
}
\label{fig:3}
\end{figure*}

Now we show results over empirical networks.
Zachary's karate-club~\cite{zachary1977information} is a network composed of 34 nodes representing members of a karate-club and 78 links representing friendship relations among them.
After a conflict arose between two important members of the club, the network broke into two communities of 17 members each, usually considered as the meta-data community structure to be contrasted against predictions.
In Fig.~\ref{fig:3}a the community structure of Zachary's network detected by the BMDL-EDM
method at $\beta=0^+$ is shown.
With exception of node 9---whose condition of outlier can be explained using information not represented in the network data~\cite{peel2017ground}---the detected four communities constitute a refinement of the community structure conveyed by the meta-data.
Similar results have been previously found by different community detection methods and,
not surprisingly, this includes the maximization of the GN modularity~\cite{brandes2008modularity,
fortunato2010community,
wu2015multi}.

Fig.~\ref{fig:3}b is analogous to Fig.~\ref{fig:3}a but for the American College football network~\cite{girvan2002community} 
This time, the best numerical results are obtained at $\beta\approx 2$.
Like for Zachary's network, the community structure detected by the BMDL-EDM
method is reasonably consistent with the community structured implied by the meta-data.
Only minor differences are found, most of them characterized by nodes 32, 42, 80, 82 and 90 which correspond to the independent teams.

\subsubsection{
\label{sec:SBP_RMDL_EDM}
The B'MDL-EDM method for community detection
and the BP algorithm
}

For simplicity, in this section we restrict our considerations to the particular case of $\beta\to 0^+$, a condition that is consistent with the choice $P(\beta)=\delta(\beta)$ (see Eq.~\ref{eq:metaRMDL_Z} and related text).
Later works may explore the more general case of $\beta>0$.

We can combine the B'MDL formalism with the family of EDMs to obtain what we called 
the B'MDL-EDM method
for community detection.
According to the formulation of section~\ref{sec:statistical_significance}, it consist in the characterization of the properties of the hyper thermodynamic system defined by the 
hyper Hamiltonian
\begin{equation}
\label{eq:meta_beta_RMDL_EDM:H:a}
H'_{G}(\mathcal{P})
:=
\mathcal{H}_{0^+,\mathcal{P}}(G)
=
\mathcal{H}^{(0)}_{\mathcal{P}}(G)
\end{equation}
where
\begin{equation}
\label{eq:meta_beta_RMDL_EDM:Z}
Z'_G(\beta')
=
\sum_{\mathcal{P}}
e^{-\beta' H'_{G}(\mathcal{P})}
\end{equation}
is the corresponding hyper partition function.
Here, summation runs over the set of all possible labellings of the nodes of $G$.
After some algebraic manipulation used to remove factors and terms constant in $\mathcal{P}$, the hyper Hamiltonian can be rewritten as
\begin{eqnarray}
\label{eq:meta_beta_RMDL_EDM:H:b}
H'_G(\mathcal{P})
&=&
\sum_{i,j\in V:i<j}
G_{ij}
\delta_{p_i p_j}
-
K_{ij}
\delta_{p_i p_j}
\end{eqnarray}
where we remind the reader that $G_{ij}$ is the adjacency matrix of a sparse-network while
$K_{ij}:=\tfrac{k_i k_j}{2}(\tfrac{1}{K-k_i}+\tfrac{1}{K-k_j})$ is the adjacency matrix of a fully connected network.
An approximate characterization of the thermodynamic properties of the system defined by  
Eqs.~\ref{eq:meta_beta_RMDL_EDM:Z}~and~\ref{eq:meta_beta_RMDL_EDM:H:b}
can be obtained by using the so called {\em Belief Propagation (BP) algorithm}~\cite{
decelle2011inference,
zhang2014scalable} in conjunction of an approximation of the hyper Hamiltonian due to Hastings~\cite{hastings2006community}, which reads
\begin{equation}
H'_G[\phi](\mathcal{P})
\approx
\sum_{i,j\in V:i<j}
G_{ij}\delta_{p_i p_j}
-
\sum_{i\in V}
K_i^{\mathrm{MF}}[\phi](p_i)
.
\end{equation}
Here, the terms
\begin{eqnarray}
K_i^{\mathrm{MF}}[\phi](p_i)
&=&
\sum_{j\in V:j\neq i}
\sum_{p_j}
\phi_j(p_j)
K_{ij}
\delta_{p_i p_j}
\\
&\approx &
\tfrac{1}{2}
\frac{k_i}{K-k_i}
\theta[\phi](p_i)
+
\tfrac{1}{2}
k_i
\eta[\phi](p_i)
\nonumber
\end{eqnarray}
are Mean Field (MF) approximations of the contribution of the terms 
$K_{ij}p_i p_j$
where
\begin{equation}
\theta[\phi](p_i)
=
\sum_{j\in V}
k_j
\phi_j(p_i)
\nonumber
\end{equation}
and
\begin{equation}
\eta[\phi](p_i)
=
\sum_{j\in V}
\frac{k_j}{K-k_j}
\phi_j(p_i)
\nonumber
.
\end{equation}
The MF terms are functionals of the so called Bethe-Peierls belief distribution 
\begin{equation}
\label{eq:belief_distribution}
\phi(\mathcal{P})
=
\prod_{ij\in E}
\phi_{ij}(p_i p_j)
\prod_{i\in V}
\phi_i^{1-k_i}(p_i)
.
\end{equation}
The quantities $\phi_i(p_i)$ and $\phi_{ij}(p_i p_j)$ represent marginal beliefs and they can be iteratively computed through the so called BP equations
\begin{eqnarray*}
n_{r\from i}^{(t+1)}(p_i)
&=&
\frac{
e^{
\beta' K_i^{\mathrm{MF}}[\phi^{(t)}](p_i)
}
}{
Z_{r\from i}^{(t+1)}
}
\prod_{j\in V_i:j\neq r}
\sum_{p_j}
e^{
-
\beta'
\delta_{p_i p_j}
}
n_{i\from j}^{(t)}
(p_j)
\\
&\approx &
\frac{1}{Z_{r\from i}^{(t+1)}}
\exp
\bigg\{
\beta'
K_i^{\mathrm{MF}}
[\phi^{(t)}](p_i)
+
\nn
&&
\;\;\;\;\;
+
\sum_{j\in V_i:j\neq r}
\ln
\bigg[
1
+
\bigg(
e^{-\beta'}
-
1    
\bigg)
n_{i\from j}^{(t)}(p_i)
\bigg]
\bigg\}
\nonumber
\end{eqnarray*}
and
\begin{eqnarray*}
\phi_i^{(t+1)}(p_i)
&\approx &
\frac{1}{
Z_i^{(t+1)}}
\exp
\bigg\{
\beta'
K_i^{\mathrm{MF}}
[\phi^{(t)}](p_i)
+
\nn
&&
\;\;\;\;\;
+
\sum_{j\in V_i}
\ln
\bigg[
1
+
\bigg(
e^{-\beta'}
-
1    
\bigg)
n_{i\from j}^{(t)}(p_i)
\bigg]
\bigg\}
\end{eqnarray*}
which also introduce the so called BP messages $n_{i\from j}(p_i)$.
When the iteration converges, the obtained fix point determines the messages, the marginal beliefs and the pairwise marginal beliefs via the equation
\begin{equation}
\phi_{ij}(p_i p_j)
=
\frac{e^{-\beta G_{ij}\delta_{p_i p_j}}}{Z_{ij}}
n_{j\from i}(p_i)
n_{i\from j}(p_j)
.
\end{equation}
Together, the marginal beliefs can be used to obtain Bethe's approximation of the hyper free energy 
\begin{equation}
\label{eq:BP_hyper_free_energy}
{F'}_{G}^{\mathrm{B}}[\phi](\beta')
=
U'_G[\phi](\beta')
-
\tfrac{1}{\beta'}
{S'}_{G}^{\mathrm{B}}[\phi](\beta')
%.
\end{equation}
where
\begin{equation}
U'_G[\phi](\beta')
=
\sum_{ij\in E}
\sum_{p_i p_j}
\phi_{ij}(p_i p_j)
\delta_{p_i p_j}
-
\sum_{i\in V}
\sum_{p_i}
\phi_i(p_i)
K_i^{\mathrm{MF}}(p_i)
\nonumber
\end{equation}
is the variational approximation of the hyper  thermodynamic energy $U'_G(\beta')$ and
\begin{eqnarray}
{S'}_{G}^{\mathrm{B}}[\phi](\beta')
&=&
-
\sum_{ij\in E}
\sum_{p_i p_j}
\phi_{ij}(p_i p_j)
\ln 
\phi_{ij}(p_i p_j)
-
\\
&&
\;\;\;\;\;
-
\sum_{i\in V}
(1-k_i)
\sum_{p_i}
\phi_i(p_i)
\ln
\phi_i(p_i)
\nonumber
\end{eqnarray}
is Bethe's approximation of the hyper thermodynamic entropy $S'_G(\beta')$.
From Eq.~\ref{eq:BP_hyper_free_energy}, a so called Bethe's approximation ${L'}_G^{\mathrm{B}}[\phi](\mathcal{P})$ of the B'MDL-EDM can be straightforwardly obtained.

The BP equations should be iterated starting from random normalized initial conditions for the beliefs and messages.
Convergence is typically easy at the high-temperature disordered phase or at any ordered phase.
However, convergence may result difficult or impossible at a {\em glassy} low temperature disordered phase~\cite{fischer1993spin,
zhang2014scalable}.
As explained in section~\ref{sec:statistical_significance},
ordered phases correspond to the existence of statistically significant model selections which, in the context of community detection via the B'MDL-EDM, this corresponds to the existence of statistically significant node partitions.

\begin{figure*}
\mbox{\includegraphics*[width=.31\linewidth]{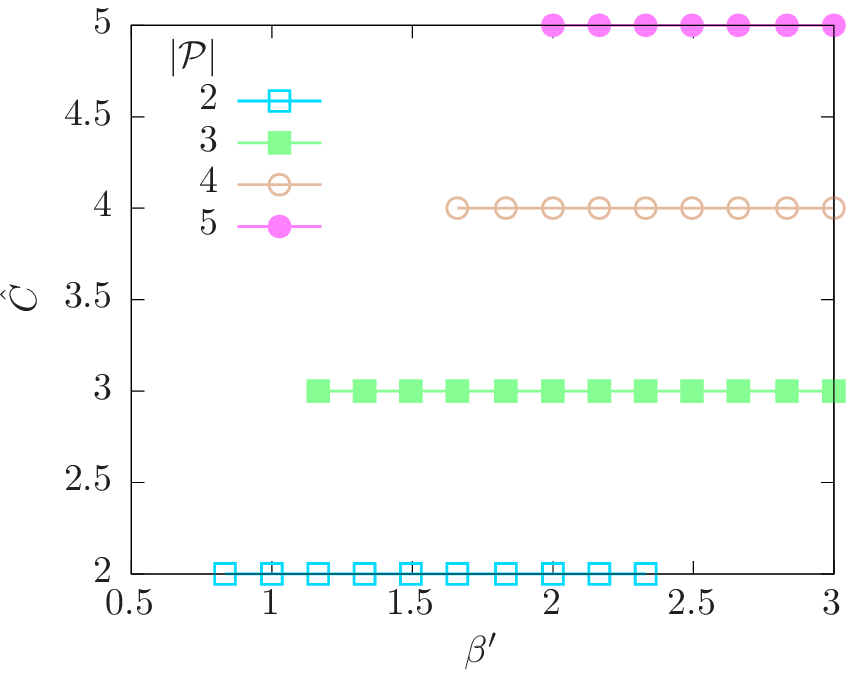}}
\hfill 
\mbox{\includegraphics*[width=.31\linewidth]{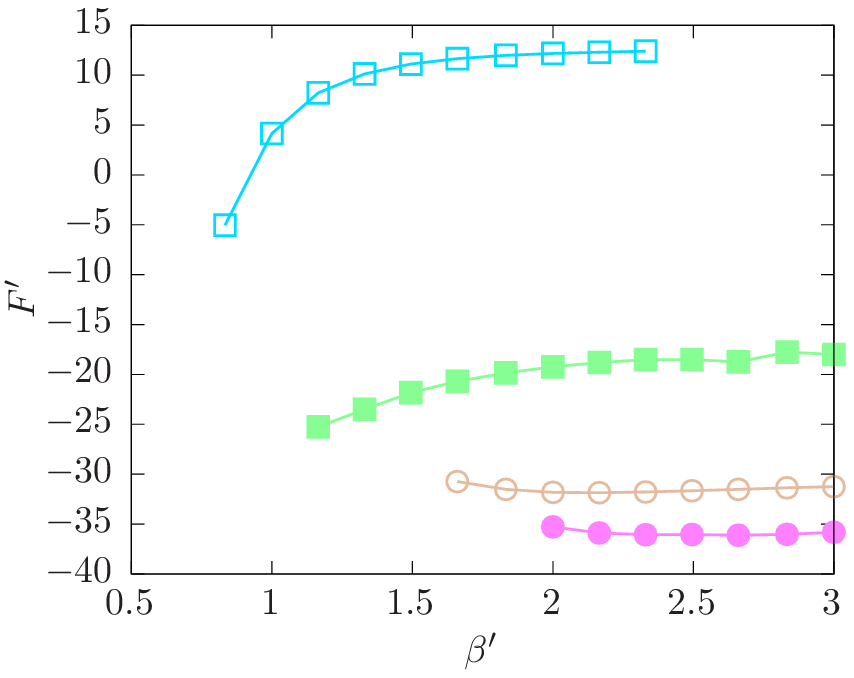}}
\hfill
\mbox{\includegraphics*[width=.31\linewidth]{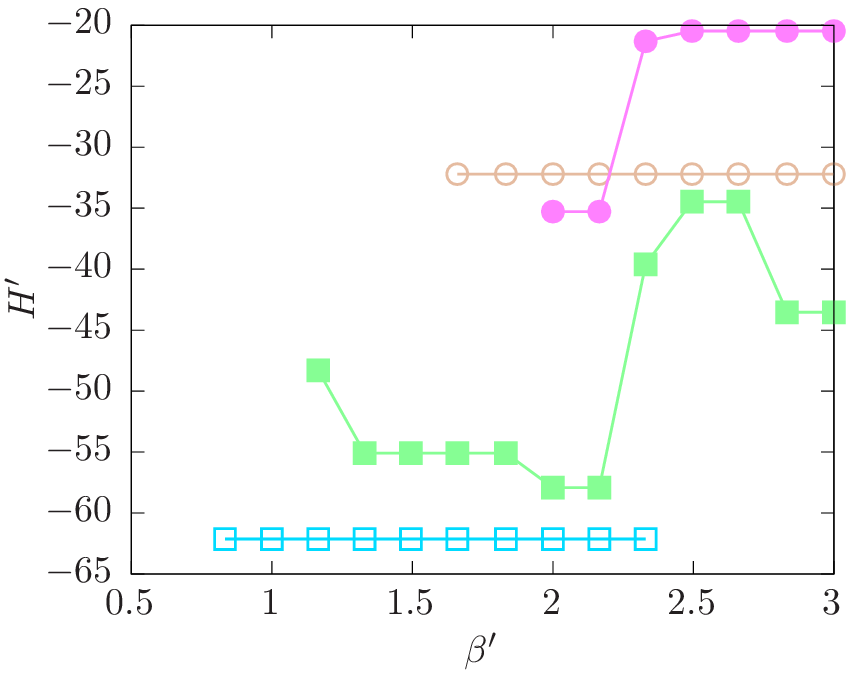}}
\put(-505,120) {a)}
\put(-335,120) {b)}
\put(-160,120) {c)}
\caption{
%(Color Online).
Estimation of the number of communities within Zachary's karate-club network and with the BP approximation of the B'MDL-EDM method for community detection.
The different curves are computed for different number of labels $|\mathcal{P}|$ and represent  functions of the hyper inverse temperature $\beta'$.
The cyan open squares corresponds to the number of communities implied by the meta-data, $C=2$.
In panel {\em a)}, the naive estimation $\hat{C}$ artificially grows with $|\mathcal{P}|$.
The analogous occur in panel {\em b)} where the hyper free energy $F'$ artificially decreases with $|\mathcal{P}|$, corresponding to a prediction where $C_{F'}\geq 5$.
In panel {\em c)}, the minimum value of the hyper Hamiltonian $H'$ occurs at the estimation $C_{H'}=2$, which is the number of communities implied by the meta-data.
}
\label{fig:4}
\end{figure*}

Once the marginals $\phi$ have been computed, the information these carry can be projected into different quantities of interest.
For example, a highly representative community structure of the network denoted by
$\hat{\mathcal{P}}$, is inferred as~\cite{zhang2014scalable}
\begin{equation}
\label{eq:estimate_hat_p}
\hat{p}_i
=
\argmax{p}\; \phi_i(p)
,
\end{equation}
from where a naive estimation
$\hat{C}:=|\mathcal{\hat{P}}|$ 
of the actual number of communities in the network, $C$, can be also obtained.
The estimation is naive because the incorrectly inferred labels $p_i$ may increase $\hat{C}$ but never decrease it.
Thus, alternative and potentially better methods for the estimation of $C$ were proposed.
For example, one way is to compute $\phi$ for increasing values of the number of labels $|\mathcal{P}|$ until the BP free energy $F'$ stops decreasing and $C_{F'}:=|\mathcal{P}|$ becomes the corresponding estimate of $C$.
Previous studies~(see Fig.~8 of~\cite{decelle2011asymptotic}) indicate that this second estimate works well on synthetic networks but gives a strange answer in Zachary's karate-club network.
This result can be understood from the information theoretic point of view defined by the B'MDL formalism.
In essence, changing the number of labels $|\mathcal{P}|$ corresponds to different choices for the null model, leading to non comparable values of the corresponding codeword-lengths B'MDL-EDM and, therefore, to non comparable free energies $F'$. 
The previous analysis suggests why a third method originally proposed by ZM in~\cite{zhang2014scalable} works better.
In fact, while the hyper entropy $S'$ may arbitrarily grow with $|\mathcal{P}|$---resulting in an arbitrarily decreasing value of $F'$---the values of $H'$ and $U'$ are bounded from below and, in consequence, the estimate $C_{H'}:=|\mathcal{\hat{P}}|$ of $C$ that is obtained by replacing the minimization of $F'$ with that of $H'(\hat{P})$ is bounded from above.
We confirmed all these observations with our 
B'MDL-EDM variant of the BP method for community detection against the planted partition network model and Zachary's karate club network.
In Fig.~\ref{fig:4}, the three different estimation of $C$ are plotted for varying  values of the hyper temperature $\beta'$ and number of labels $|\mathcal{P}|$.
We show results for Zachary's karate-club network only, since it corresponds to the most interesting case.
In panel~\ref{fig:4}a, the estimation $\hat{C}$ arbitrarily grows with $|\mathcal{P}|$ far beyond the number of communities suggested by the meta-data (cyan open squares) and for all values of $\beta'$.
In panel~\ref{fig:4}b, the curves for the hyper free energy $F'$ monotonously decrease with growing $|\mathcal{P}|$, resulting in an over-estimation $C_{F'}$ of $C$.
Finally, in panel~\ref{fig:4}c the hyper Hamiltonian $H'$ reaches its minimum at the number of communities implied by the meta-data and thus resulting in an accurate prediction.

\section{Conclusion}
\label{sec:conclusions}

In this work we introduce the Boltzmannian Minimum Description Length (BMDL),
a framework for statistical modeling that is grounded on the principle of MDL of J. Rissanen~\cite{barron1998minimum,
grunwald2007minimum}, whose parametric complexity
can be formally related to an artificial
statistical mechanical systems in thermodynamic equilibrium.
Taking advantage of the rich theoretical and technical background of statistical mechanics, we leverage on the information theoretic aspects of statistical modeling to show the crucial role that phase transitions play in the BMDL formalism. More specifically, to justify: {\em i)} a high-temperature series expansion to compute systematic approximations of the BMDL codeword-length when it is used to model data and {\em ii)} the association of ordered phases with statistical significant model selections when the BMDL is used to {\em model models}.
Our framework presents several advantages over previously introduced ones.
For example, compared to the Bayesian approach, the BMDL overrides the need of priors in favor of more comprehensible null models whose choice can be easily driven, not only by mathematical convenience, but also by the research question.
Moreover, the BMDL is a framework connecting information theory with both, the frequentist and Bayesian approaches to statistics, which remain invariant under the transformations of the model's parameters~\cite{hansen2001model}---a possibility also valid within the Bayesian approach, but only with Jeffrey's priors.

We illustrate the power of the introduced formalism in a couple of practical examples.
Firstly, and briefly, we show how the high-temperature series expansion or the BMDL can be used to characterize and treat the divergences emerging in calculation of the RMDL that occur for certainly simple statistical problems.
Secondly, we extensively test the formalism against the challenging problem of community detection in complex networks.
For this purpose, we combine the BMDL with a parameterized family of statistical models, from where a principled derivation of the Girvan-Newman (GN) modularity~\cite{girvan2002community,
newman2016equivalence} can be obtained as the first order approximation of the high-temperature series expansion.
Moreover, when the BMDL framework is used to model models, we show how to derive the Belief Propagation method for community detection of Zhang-Moore (ZM)~\cite{zhang2014scalable}.
We study the derived community detection methods by mean of analytical considerations and numerical experiments on synthetic and empirical networks to find: {\em iii)} that the correction terms to the GN modularity introduced by the high-temperature series expansion improve the performance of community detection, {\em iv)} an information theoretic justification of why the ZM criteria for the inference of the number of network communities is better than other proposed alternatives and {\em v)} in agreement with recent findings using a large set of empirical networks and community detection methods~\cite{ghasemian2018evaluating}, our results reinforce the idea that optimization based community detection methods tend to over-estimate the number of communities, while integral based community detection methods behave more conservatively.
In this regard, our formalism is advantageous since, at least in principle, it provides a way to interpolate between these two cases for almost any score-based community method that can be proposed.

Finally, let us mention a few possibilities of the numerous opportunities our present contribution opens for future work.
In the general context of statistical modeling, there is plenty of work to be done to test the performance, benefits and drawbacks of the BMDL formalism.
This can be studied from both, a theoretical perspective and a practical perspective.
For example, the study of non-uniform null models---which can be introduced adding an extra and common term to the Hamiltonians, $\beta H_m(x) + H_0(x)$---is an interesting road to follow.
Also, a more profound and general  treatment of the divergences sometimes displayed by the RMDL, or the consideration of other statistical mechanical ensembles~\cite{squartini2015breaking}are among other options.
On the other hand, in the particular context of community detection in complex networks, it would be useful to consider other families of models generalizing the role of EDM model and  the corresponding null model to the case of weighted, directed, disassortative, temporal, multi-layer and hierarchical community structures~\cite{mucha2010community,
holme2012temporal,
perotti2015hierarchical,
yang2017hierarchical} among others, or even to characterize other kind of network properties for the generalization of null models for network reconstruction~\cite{cimini2015systemic,
squartini2017enhanced}.

%\section{Acknowledgments}
%\label{sec:acknowledgments}

\begin{acknowledgments}
We thank O.V. Billoni, A.L. Schaigorodsky, N. Almeira and F. Saracco for useful discussions.
JIP acknowledges financial support from PIP CONICET nr. 112 201501 00285 and institutional support from IFEG-CONICET and FaMAF-UNC.
JIP and GC acknowledge support from FP7-ICT project MULTIPLEX nr. 317532, FP7-ICT project SIMPOL nr. 610704, and Horizon 2020 project DOLFINS nr. 640772.
CJT acknowledges financial support of the University Research Priority Program on Social Networks, University of Z\"urich.
AC acknowledges support from Grant No. IIS-1452718 from the National Science Foundation.
\end{acknowledgments}

%%%%%%%%%%%%%%%%%%%%%%%%%%
%%\begin{thebibliography}{99}
%%\end{thebibliography}
%
%\bibliographystyle{apsrev4-1}
%\bibliography{ref}


\end{document}